\documentclass[prl,twocolumn,showpacs,superscriptaddress]{revtex4}

\usepackage{mathrsfs,mathtools,bbold}
\usepackage{amsmath,amssymb}
\usepackage{verbatim}
\usepackage{graphicx}
\usepackage{ccaption}
\usepackage[dvipsnames]{xcolor}
\usepackage[colorlinks=true,linkcolor=blue,citecolor=magenta,urlcolor=MidnightBlue,filecolor=black]{hyperref}

\DeclareFontFamily{OT1}{rsfs}{}
\DeclareFontShape{OT1}{rsfs}{m}{n}{ <-7> rsfs5 <7-10> rsfs7 <10->rsfs10}{} 
\DeclareMathAlphabet{\mycal}{OT1}{rsfs}{m}{n}

\newcommand{\e}{\epsilon}

\newcommand{\C}{\tilde{C}}
\renewcommand{\L}{{\mathcal{L}}}
\newcommand{\bL}{\bar{{\mathcal{L}}}}

\newcommand{\be}[1]{ \begin{equation}\label{#1} }
\newcommand{\ee}{\end{equation}}
\newcommand{\bea}[1]{\begin{eqnarray}\label{#1} }
\newcommand{\eea}{\end{eqnarray}}
\newcommand{\p}{\partial}
\newcommand{\non}{\nonumber}

\newcommand{\refb}[1]{(\ref{#1})}

\renewcommand{\>}{\rangle}

\renewcommand{\a}{\alpha}
\newcommand{\ta}{\tilde{\a}}
\renewcommand{\b}{\beta}
\renewcommand{\t}{\tau}
\newcommand{\s}{\sigma}
\newcommand{\za}{|0\rangle_\a}
\newcommand{\zc}{|0\rangle_c}

\begin{document}
 
\title{Rindler on the Worldsheet}

\author{Arjun Bagchi}
\email{abagchi@iitk.ac.in}
\affiliation{Indian Institute of Technology, Kanpur 208016, India.}

\author{Aritra Banerjee}
\email{aritra.banerjee@apctp.org}
\affiliation{Asia Pacific Center for Theoretical Physics, Postech, Pohang 37673, Korea.}

\author{Shankhadeep Chakrabortty.}
\email{s.chakrabortty@iitrpr.ac.in}
\affiliation{Indian Institute of Technology Ropar, Rupnagar, Punjab 140001, India.}

\begin{abstract} 
We construct the tensionless limit of bosonic string theory in terms of a family of worldsheets with increasing acceleration and show that the null string emerges in the limit of infinite acceleration when the Rindler horizon is hit. We discover a novel phenomenon we call null string complementarity, which gives two distinct observer dependent pictures of the emergence of open string physics from closed strings in the tensionless limit. The closed string vacuum as observed by the inertial worldsheet turns into a D-instanton in the tensionless limit, while in the complementary picture from the accelerated worldsheet one sees the emergence of a D-25 brane. We finally discuss approaching the Rindler horizon through time evolution at constant acceleration and also show how an open string picture arises very naturally.  
\end{abstract}


\maketitle

\noindent {\em{\underline{Introduction}}}.  
The study of physics in accelerated frames of reference is an intriguing and fruitful venture. Accelerated observers in Minkowski spacetimes view the Rindler metric and experience a horizon. The entire Minkowski diamond is not accessible to them anymore and in this frame of  reference, physics is thermal. Due to the unavailability of information from beyond the Rindler horizon, the density matrix of this accelerated observer is related to that of an inertial Minkowski observer by partially tracing over the inaccessible degrees of freedom. All of this is of course well understood and is very useful for black hole physics since the near-horizon limit of a black hole typically yields a Rindler spacetime. Our discussions in this paper are rather unique. We want to understand aspects of Rindler physics {\em on the worldsheet of a closed string}. Our motivation for doing so is also rather novel, as we elaborate below. 

String theory is currently the most promising of avenues for formulating a theory of quantum gravity. One of its primary endearing features is the paucity of tuneable parameters. The free theory has only one, the length of the fundamental string ($\ell_s$). $\ell_s\to0$ reduces the string to a point particle and the theory to the well understood Einstein's theory of general relativity. In this work, we are interested by the other extreme limit, where $\ell_s\to\infty$ \cite{Schild:1976vq}. This bizarre limit corresponds to the ultra-stringy regime of string theory that is very different to Einstein's theory. The quantum version of this theory would be ``very stringy" quantum gravity. This limit also takes the tension of the fundamental string to zero. Our objective in this paper is to {\em formulate the decreasing string tension in terms of accelerated string worldsheets}.

The tensionless string is a null string with a degenerate worldsheet metric. We will show in this paper that the worldsheet analogue of hitting the Rindler horizon leads to the formation of the null string. Rindler observers can approach the horizon in two distinct ways: time evolution at a fixed acceleration or evolution in acceleration at a constant time. The limit from the tensile to the tensionless string can also be formulated in terms of increasingly accelerated worldsheets or time-evolution on a constant acceleration worldsheet. We shall follow both routes with interesting consequences, the most intriguing amongst which is what we call {\em null string complementarity}. Depending on whether the observer sits on an inertial worldsheet and observes an accelerated one, or vice-versa, she sees the emergence of different complementary boundary states as the closed string becomes tensionless. The inertial worldsheet sees the accelerated closed string vacuum evolve into a spacetime point, a D-instanton, while the infinitely accelerated observer sees the inertial vacuum grow into a spacefilling D-25 brane. No one observer has access to both pictures.

\smallskip

\noindent {\em{\underline{Rindler Physics}}}. Accelerated observers in Minkowski spacetimes moving on trajectories $x^2 - t^2 = \kappa^{-2}$ ($\kappa$ is proper acceleration) describe Rindler space with metric 
\be{}
ds_{R}^2 = e^{2a\xi}(-d\eta^2+d\xi^2).
\ee
Minkowski and Rindler spacetimes are linked by: 
\be{Rmap}
t = \frac{1}{a}e^{a\xi}\sinh a\eta,~~ x= \frac{1}{a}e^{a\xi}\cosh a\eta.
\ee 
where $\kappa =a e^{-a\xi}$, $a$ is the redefined acceleration. \refb{Rmap} is for the right Rindler wedge only ($R: |t|< x, x>0$). There is an equivalent left wedge ($L: |t|< x, x<0$), for which Eqs \refb{Rmap} pick up negative signs. 
\begin{figure}[h]
\centering
  \includegraphics[width=50mm]{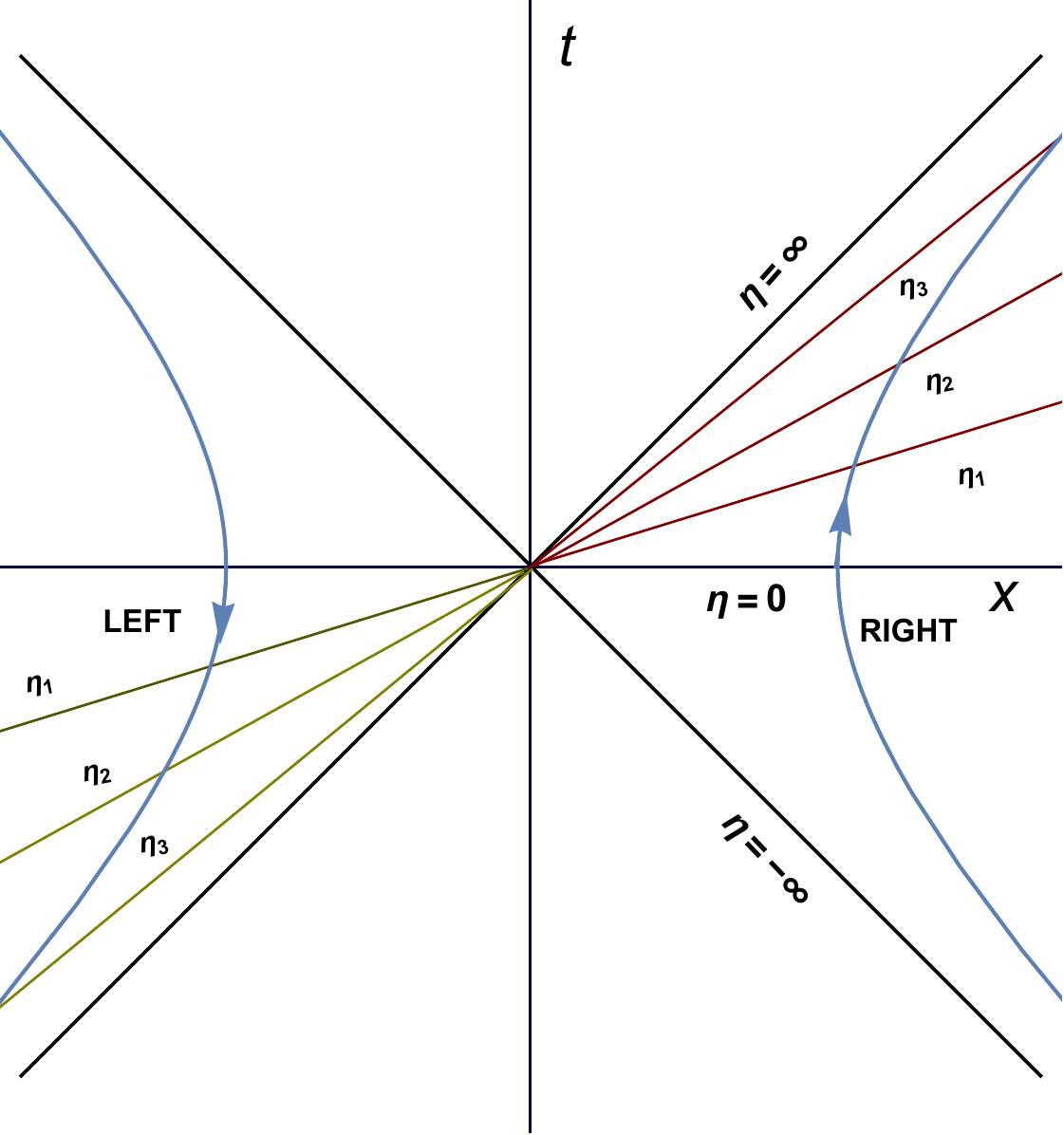}
  \caption{Equal time slices in Rindler spacetimes}
  \label{fig1}
\end{figure}

We now consider a massless scalar field theory in Rindler spacetime \cite{Birrell:1982ix}. Since flat and Rindler backgrounds are conformally related, the equations of motion (EOM) for the field are same: 
\be{}
\Box_{t,x}\phi =0=\Box_{\eta,\xi}\phi.
\ee
For the Minkowski solution, now defined on a cylinder $(\s,\t)$, the wave equation is solved by
\bea{tensilemode}
  \phi (\s, \t) &=&\phi_0+\sqrt{2\a'}\a_0\t \\ && +i\sqrt{\frac{\a'}{2}}\sum_n\Big[\frac{\a_n}{n} e^{-in(\t+\s)}+\frac{\tilde{\a}_n}{n} e^{-in(\t-\s)}\Big] \non
   \eea
where oscillators satisfy $[\a_n,\a_m]=n\delta_{n+m}$ and annihilate the Minkowski vacuum $|0\>_M$. To match with usual scalar field modes, we write \refb{tensilemode}:
\bea{}
\phi (\s, \t) &=& \phi_0+\sqrt{2 \alpha'} \alpha_0\tau \\ && + \sqrt{2 \pi \alpha'}\sum_{n>0}[\a_n u_n+\a_{-n} u_n^*+\ta_n \tilde{u}_n+\ta_{-n} \tilde{u}_n^*] \non
\eea
where $u_n = \frac{i e^{-in(\tau+\sigma)}}{\sqrt{4\pi}n},~~ \tilde{u}_n=\frac{i e^{-in(\tau-\sigma)}}{\sqrt{4\pi}n}$.
Similarly, we write the Rindler mode expansion as 
\bea{}
\phi(\xi, \eta) &=& \phi_0+\sqrt{ 2 \alpha'}\beta_0\xi  \\&&+ \sqrt{2 \pi\alpha'}\sum_{n>0}[\b_n U_n+\b_{-n} U_n^*+\tilde{\b}_n \tilde{U}_n+\tilde{\b}_{-n} \tilde{U}_n^*] \non
\eea
where the mode functions are now defined as
\be{}
U_n = \frac{i e^{-in(\xi+\eta)}}{\sqrt{4\pi}n},~~ \tilde{U}_n=\frac{i e^{-in(\xi-\eta)}}{\sqrt{4\pi}n}.
\ee
The oscillators $(\b,\tilde\b)$ now act on a new vacuum $|0\rangle_R$. Importantly, $U_n$ is only defined in the L wedge (hence called $U_n^{(L)}$) and $\tilde{U}_n$ only in the R wedge ($U_n^{(R)}$), unlike flat space. For continuing between wedges one needs to define smearing functions. The combinations analytically continued in both wedges take the form \cite{Unruh:1976db}: 
\be{}
U_n^{(R)} - e^{-\frac{\pi n}{a}}U_{-n}^{(L)*}, \quad U_{-n}^{(R)*} - e^{\frac{\pi n}{a}}U_{n}^{(L)}.
\ee
Using these combinations of modes lead us to Bogoliubov transformations between the two sets of oscillators in Rindler and Minkowski space with a form \cite{Unruh:1976db}: 
\begin{subequations}\label{R2M}
\bea{}
\b_{n} &=& \frac{e^{\pi n/2a}}{\sqrt{2\sinh\frac{\pi n}{a}}}~\a_{n}-\frac{e^{-\pi n /2a}}{\sqrt{2\sinh\frac{\pi n}{a}}}~\tilde{\a}_{-n},\\
\tilde{\b}_{n}&=& -\frac{e^{-\pi n/2a}}{\sqrt{2\sinh\frac{\pi n}{a}}}~\a_{-n}+\frac{e^{\pi n/2a}}{\sqrt{2\sinh\frac{\pi n}{a}}}~\tilde{\a}_{n}.
\eea
\end{subequations}
Eq. \refb{R2M} forms the backbone of our analysis in this paper. 

\smallskip

\noindent {\em{\underline{Intrinsic look at Tensionless strings}}}.
The starting point of our recapitulation is the action \cite{Isberg:1993av}
\be{LST}
S_{\text{ILST}} = \int d^2 \xi \, V^\a V^\b \p_\a X^\mu \p_\b X^\nu \eta_{\mu \nu}
\ee
\refb{LST} can be obtained from the Polyakov action for the bosonic string as tension $T \to0$ \cite{Isberg:1993av}, where vector densities $V^\a$ replace the degenerate worldsheet metric. Like in tensile string theory, \refb{LST} enjoys worldsheet diffeomorphism symmetry and needs to be gauge fixed. In $V^\alpha = (1, 0)$ gauge, the residual symmetry is
\bea{bms}
&& [L_n, L_m] = (n-m) L_{n+m} + c_L\delta_{n+m,0} (n^3-n), \cr
&& [L_n, M_m] = (n-m)M_{n+m} + c_M\delta_{n+m,0} (n^3-n), \cr 
&& [M_n, M_m]=0.
\eea 
This is the BMS$_3$ algebra (here with $c_L=c_M=0$), which also arises as the asymptotic symmetries of 3d flat spacetimes at its null boundary \cite{Barnich:2006av}, and has been used to construct a notion of Minkowskian holography following \cite{Bagchi:2010eg}. For tensionless strings BMS$_3$ replaces the two copies of the Virasoro algebra that dictate the construction of tensile strings \cite{Bagchi:2013bga}. 

In $V^\alpha = (1, 0)$ gauge, the EOM of $V^\a$ give constraints while the EOM for $X$ takes a simple form:
\be{xeom}
\ddot{X}^\mu=0; \quad {\text{Constraints:}} \ \dot{X}\cdot X'=0, \, \dot{X}^2=0. 
\ee
With closed string boundary conditions $X^\mu(\tau,\sigma)=X^\mu(\tau,\sigma+2\pi)$, EOM can solved by \cite{Bagchi:2015nca}
\be{mode} 
X^{\mu}(\sigma,\tau)=x^{\mu}+\sqrt{\frac{c'}{2}}B^{\mu}_0\tau+i\sqrt{\frac{c'}{2}}\sum_{n\neq0}\frac{1}{n} \left(A^{\mu}_n-in\tau B^{\mu}_n \right)e^{-in\sigma}. 
\ee
In the above, $c'$ is a length-scale introduced for dimensional consistency. This expansion also leads to the constraints in the form
\be{lmab} 
L_n= \frac{1}{2} \sum_{m} A_{- m}\cdot B_{m+n}, \ M_n= \frac{1}{2} \sum_{m} B_{-m}\cdot B_{m+n}. 
\ee
$A,B$ are not the usual harmonic oscillator modes:
\bea{AB} 
[A_m,A_n]= [B_m,B_n]=0; [A_m,B_n]= 2 m\delta_{m+n}.
\eea
The algebra of constraints leads to the BMS$_3$ algebra as before. We transform $(A,B)$ into a harmonic oscillator basis: 
\be{CC}
2C^{\mu}_n = ({A}^{\mu}_n+B^{\mu}_{n}), \quad 2\C^{\mu}_n = (-{A}^{\mu}_{-n}+B^{\mu}_{-n}).
\ee
$C, \C$ now have canonical commutation relations analogous to tensile $\a$ oscillators. Mode expansion in terms of these $C$ modes reads \cite{Bagchi:2020fpr} 
\bea{cexpansion}
X^\mu(\s,\t)&=&x^\mu+2\sqrt{\frac{c'}{2}}C^\mu_0\t +i\sqrt{\frac{c'}{2}} \\ && {\hspace{-1cm}} \times \sum_{n\neq 0} \frac{1}{n}\left[(C^\mu_n-\C^\mu_{-n})-in\t (C^\mu_n+\C^\mu_{-n})\right]e^{-in\s} \non
\eea
with zero modes $C^\mu_0=\C^\mu_0=\sqrt{\frac{c'}{2}}k^\mu.$  

\smallskip

\noindent {\em{\underline{Tensionless strings as a Carrollian limit}}}.
In the discussion above, the string tension was put exactly to zero. Now we describe a limiting procedure on the worldsheet coordinates that takes the tension to zero \cite{Bagchi:2013bga,Bagchi:2015nca}:  
\be{URlim}
\s \to \s, \ \t \to \e \t, \ \a'\to c'/\e, \ \e \to 0.
\ee
This sends the worldsheet speed of light to zero and is called an Ultra-Relativistic (UR) or a Carrollian limit. In this limit, the worldsheet becomes a 2d Carrollian manifold \cite{Duval:2014uoa, Duval:2014lpa}, with a degenerate metric which is the defining feature of a tensionless or a null string. The worldsheet symmetry generators contract 
\be{vir2bms}
L_n= \L_n - \bL_{-n}, \ M_n = \e(\L_n + \bL_{-n}),
\ee
(here $\L_n, \bL_n$  generate the tensile Virasoro algebra) and close to form BMS$_3$. Comparing tensile modes (analogue of \refb{tensilemode}) and the tensionless expansions \refb{mode} we get
\be{ABa}
A_n^{\mu} = \frac{1}{\sqrt{\e}} \left( \a_n^\mu - \tilde{\a}_{-n}^\mu \right), \quad B_n^{\mu} = {\sqrt{\e}} \left( \a_n^\mu + \tilde{\a}_{-n}^\mu \right). 
\ee
All classical physics of the tensionless string can be reproduced by following this UR limit. The Carrollian limit is usual perceived as a limit on velocities and hence an infinite boost. The tensionless limit is thus an infinite boost on the Riemannian worldsheet of a tensile string that turns it into a degenerate Carrollian worldsheet. 

Interestingly, switching to the language of  $C$ oscillators, we find the emergence of a worldsheet Bogoliubov transformation:
\bea{c1}
C^\mu_n =\frac{1}{2} \left( \sqrt{\e} + \frac{1}{\sqrt{\e}} \right) \a^\mu_n+\frac{1}{2} \left( \sqrt{\e} - \frac{1}{\sqrt{\e}} \right) \ta^\mu_{-n}, \cr 
\C^\mu_n =\frac{1}{2} \left( \sqrt{\e} - \frac{1}{\sqrt{\e}} \right) \a^\mu_{-n}+\frac{1}{2} \left( \sqrt{\e} + \frac{1}{\sqrt{\e}} \right) \ta^\mu_{n}, 
\eea
Although the $C$ oscillators have been defined near $\e\to0$, curiously at $\e=1$ they reduce to the $\a$ oscillators. This encourages us to define a flow valid throughout the parameter space $\e \in [0,1]$. For any evolving oscillator $C(\e)$ interpolating between $\a$ for $\e=1$ and \refb{c1} near $\e\to0$, the vacua defined by the flow $|0(\e)\>$ changes continuously with $\e$:
\be{oe}
|0(\e)\>: \, C_n(\e) |0(\e)\> = \C_n(\e) |0(\e)\> = 0, \, \forall n>0.
\ee
An evolution in boost can only lead to changes in physics (e.g. change in vacuum structure, changes in spectrum) in the limit of infinite boosts. This $\e$ evolution changes the vacuum continuously and hence cannot be thought of as an evolution in boosts. As we will now see, this evolution in parameter-space is very naturally explained by accelerating string worldsheets. The identification \refb{c1} stays valid near $\e\to0$, while a map for the whole of the parameter space emerges through acceleration. 

\smallskip

\noindent {\em{\underline{Reaching the Horizon I: Evolving in acceleration}}}. We now build the string equivalent of a Rindler observer approaching the Rindler horizon by considering a family of worldsheets with increasing values of acceleration. In the limit of large acceleration, the Bogoliubov coefficients \refb{R2M} become:
\bea{binfty}
\b_{n}^{\infty} = \frac{1}{2}\left(\sqrt{\frac{\pi n}{2a}}+\sqrt{\frac{2a}{\pi n}}  \right) \a_{n}+\frac{1}{2}\left(\sqrt{\frac{\pi n}{2a}}-\sqrt{\frac{2a}{\pi n}}\right) \tilde{\a}_{-n}\cr
\tilde{\b}_{n}^{\infty}=\frac{1}{2}\left(\sqrt{\frac{\pi n}{2a}}-\sqrt{\frac{2a}{\pi n}}\right)\a_{-n}+ \frac{1}{2}\left(\sqrt{\frac{2a}{\pi n}}+\sqrt{\frac{\pi n}{2a}}\right)\tilde{\a}_{n}.\cr
\eea
This limit takes us very near the lightcone. Comparing \refb{binfty} with \refb{c1}, we see that we can make the identification 
\be{atoe}
C_n = \b_{n}^{\infty},  \ \C_n= \tilde{\b}_{n}^{\infty}, \ \e = \frac{\pi n}{2a}
\ee
The limit of zero tension is thus the limit of infinite acceleration
\be{}
\e\to 0 \Rightarrow a \to \infty
\ee
This is because the equivalence has to hold for all $n$. {\footnote{The above expressions of course don't hold for $n=0$ since the zero modes are equal \refb{cexpansion}.}} 

The evolution in parameter space alluded to earlier is thus clearly an evolution in terms of accelerated worldsheets defined for all values of acceleration. This picture of a family of accelerated worldsheets ties in nicely with our earlier description of the UR limit as the limit of infinite boost and the limit of infinite acceleration both land up on the horizon of Rindler spacetime, which in the string analogue is equivalent to the null string. 

The flow in acceleration is a flow from the tensile to the tensionless string. Hence {\em increasing acceleration amounts to decreasing string tension}, with $a=0$ being the tensile theory and $a\to \infty$ the tensionless null string. This flow is now described for all values of acceleration giving us a complete interpolating solution. The string oscillator construction through the flow is described in terms of the interpolating $\b$ oscillators, defined in \refb{R2M}. At $a=0$, from \refb{R2M}, we see that $\b$ reduces to tensile string $\a$ oscillators. For intermediate values of $a$, i.e. $0<a<\infty$, we have the $\b$ oscillators. Very near the lightcone, as $a\to \infty$, the $\b$ oscillators take the form of the tensionless $C$ oscillators.
\begin{subequations}
\bea{}
a=0 &:& \quad \{ \b_n, \tilde{\b}_n \} \to \{ \a_n, \tilde{\a}_n \} \\
0<a<\infty &:& \quad \{ \b_n (a) , \tilde{\b}_n(a) \} \\
a\to \infty &:& \quad \{ \b_n, \tilde{\b}_n \} \to \{C_n, \C_n\}
\eea
\end{subequations}
We will now use these oscillators to understand the evolution of vacuum structure of the string.

\smallskip

\noindent {\em{\underline{Structure of accelerated vacua}}}. The vacuum conditions on these accelerated worldsheets are: 
\bea{bcond}
&& \b^\mu_k|0(a)\rangle = (\alpha^\mu_k +\tanh \theta_k\ \tilde{\alpha}^\mu_{-k})|0(a)\rangle=0,\  k>0; \nonumber \\ 
&& \tilde{\b}^\mu_k|0(a)\rangle = (\tilde{\alpha}^\mu_k+\tanh\theta_k\ \alpha^\mu_{-k})|0(a)\rangle=0.
\eea 
where $\tanh{\theta_k}= -\exp{(-\frac{\pi k}{a})}$. Written in terms of the tensile closed string vacuum $|0\>_\a$, the accelerated vacuum is a squeezed state: 
\be{vacrltn}
|0(a)\rangle= \prod_{k=1}^\infty \frac{1}{\cosh\theta_k}\exp\left[-\frac{\tanh{\theta_k}}{k}\ \a_k^\dagger\cdot\tilde{\a}_k^\dagger\right]|0\rangle_\alpha.
\ee
As $a\to\infty$, the map to $\e$ \refb{atoe} emerges. We find $\tanh \theta = \frac{\e-1}{\e+1}$, so $\tanh\theta\to -1$ as $\e\to 0$, and the resulting limiting vacuum state $\zc$ is
\bea{zc2za}
\zc = \lim_{a\to\infty} |0(a)\rangle = \frac{1}{\mathcal{N}} \prod_{n=1}^\infty \exp\left[\frac{1}{n} \a_n^\dagger\cdot\tilde{\a}_n^\dagger\right]|0\rangle_\alpha. 
\eea
Here $\mathcal{N}$ is a normalisation constant. Relations between $C$ and $\a$ are invertible and $\za$ can be expressed in terms of $|0(a)\>$. This would be the string equivalent of a Rindler observer looking at her Minkowski counterpart. As $a\to\infty$, we find:
\be{za2zc}
|0\rangle_\a = \frac{1}{\mathcal{N'}} \prod_{n=1}^\infty \exp\left[- \frac{1}{n} C_n^\dagger\cdot\tilde{C}_n^\dagger\right]\zc. 
\ee

\begin{figure}[h]
\centering
  \includegraphics[width=90mm]{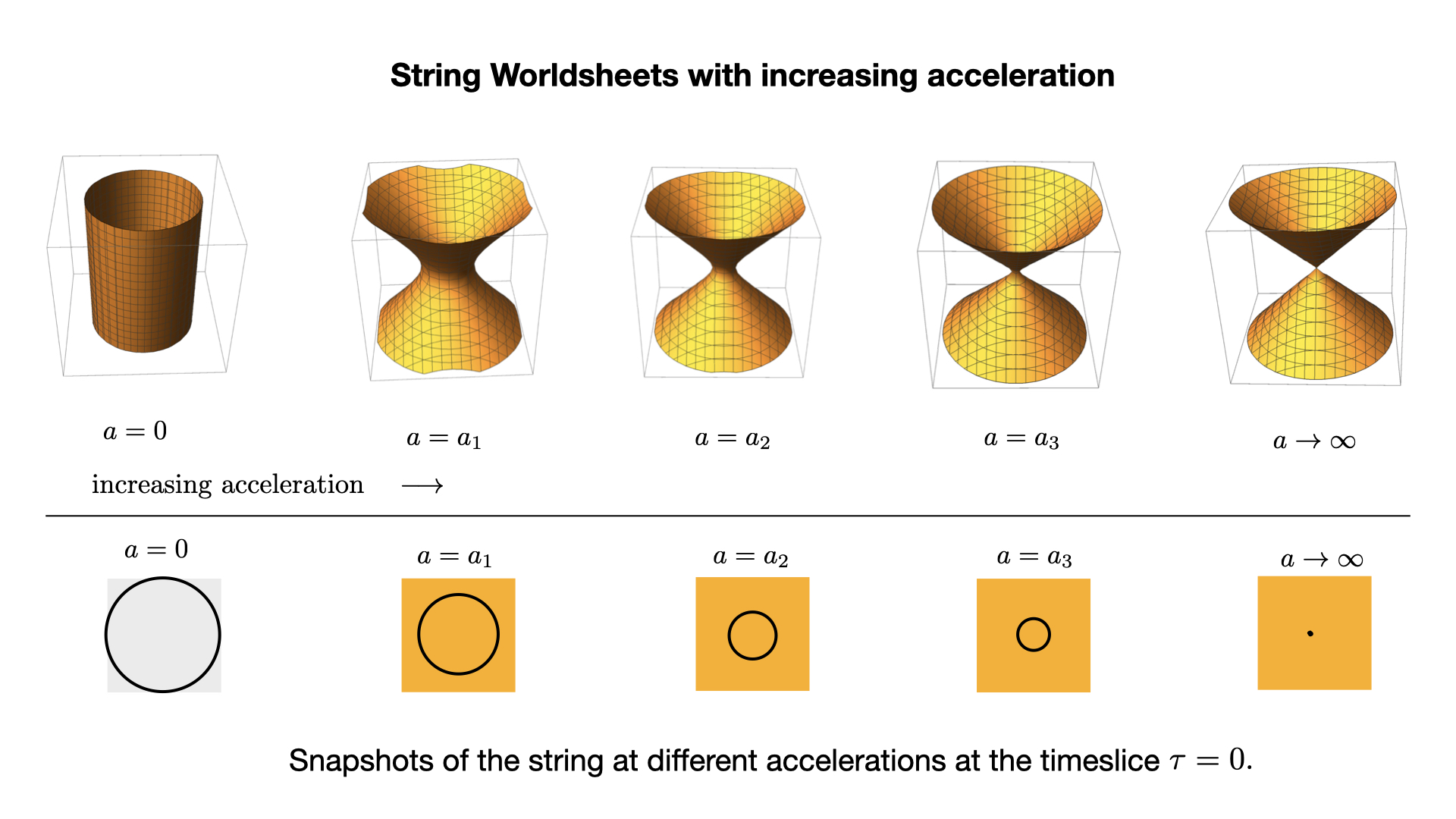}
  \caption{ Increasing accelerated worldsheets. }
  \label{fig2}
\end{figure}

\smallskip

\noindent {\em{\underline{Boundary states and Null string complementarity}}}. A tensile closed string field $X(\s,\t)$ which maps the worldsheet to spacetime is given by a mode expansion analogous to \refb{tensilemode}. D-branes arise as boundary states on the closed string worldsheet CFT. For a boundary located at $\t=0$ on the worldsheet, the possibilities are \cite{Blumenhagen:2009zz}:
\begin{subequations}\label{bst}
\bea{}
\text{N:} \ \p_\s X(\s, \t) |B_1\> = 0 \equiv (\a_n + \ta_{-n}) |B_1\> = 0. \\
\text{D:} \ \p_\t X(\s, \t) |B_2\> = 0 \equiv (\a_n - \ta_{-n}) |B_2\> = 0.
\eea
\end{subequations}
where N and D stand for Neumann and Dirichlet conditions respectively. 
This can be solved explicitly to obtain: 
\begin{subequations}
\bea{}
\hspace{-0.5cm} \text{N:} \ |B_1\> &=& {\mathcal{N}_1} \prod_{n=1}^\infty \exp\left[-\frac{\a_n^\dagger\cdot\tilde{\a}_n^\dagger}{n} \right]|0\rangle_\alpha, \\
 \text{D:} \ |B_2\> &=& {\mathcal{N}_2} \prod_{n=1}^\infty \exp\left[\frac{\a_n^\dagger\cdot\tilde{\a}_n^\dagger}{n} \right]|0\rangle_\alpha.
\eea
\end{subequations}
We see that in terms of $\a$ oscillators and the $\a$ vacuum \refb{zc2za}, $\zc$ is a {\em Dirichlet boundary state} in all directions, while in terms of the $C$ oscillators and the $C$ vacuum \refb{za2zc}, $\za$ is a {\em Neumann boundary state} (again in all directions). Thus an open string description emerges from the closed string vacuum as tension goes to zero. This can be further clarified by looking at the oscillators. Using \refb{c1} the conditions for the $C$ vacuum \refb{oe} translates into 
\bea{}
\left[\left( \sqrt{\e} + \frac{1}{\sqrt{\e}} \right)\frac{\a^\mu_n}{2} + \left( \sqrt{\e} - \frac{1}{\sqrt{\e}} \right) \frac{\ta^\mu_{-n}}{2}\right]  |0\>_c = 0 \non 
\eea
In the strict limit $\e\to0$, we end up with 
\be{}
(\a^\mu_n - \ta^\mu_{-n}) |0\>_c=  0. 
\ee
In terms of the usual string vacuum $\za$, this zero tension groundstate $\zc$ is thus a D-instanton, which is a Dirichlet boundary state in all spacetime directions \cite{Polchinski:1994fq}. An analogous calculation yields
\be{}
(C^\mu_n + \C^\mu_{-n}) \za = 0
\ee
Thus, from the point of view of the $C$ observer, the tensile string vacuum develops into a D-25 brane. 

We now physically describe this process, the one we will call the {\em null string complementarity}. For an observer in tensile string theory with vacuum $|0\>_\a$ looking at strings with decreasing tension, the completely tensionless string appears as a spacetime point, a D-instanton. There is a complementary point of view of accelerated vacua looking at $|0\>_\a$ vacuum. To the observers in this continuous 1-parameter family of vacua, the usual closed string looks more and more distorted and ultimately the tensionless observer looking at the usual string sees a spacefilling D-25 brane \cite{Bagchi:2019cay}. This is a closed to open string transition. The complementary picture of the formation of the D-instanton fits in rather wonderfully with Rindler worldsheets. This is a ``dual" picture of the formation of an open string from a closed string in the tensionless limit, as seen by observer sitting in the $\a$-vacuum. 

We pictorially depict the above process in Fig \ref{fig2}. The ``inertial" closed string worldsheet is the cylinder on the extreme left with acceleration $a=0$. For increasing accelerations $a_i$ $(a_1<a_2<a_3)$, the worldsheet can be given by increasingly distorted hyperboloids. Ultimately, at $a\to\infty$, the worldsheet becomes the lightcone. The boundary states in \refb{bst} are defined at $\t=0$, hence to understand their formation we consider the $\t=0$ cross-sections depicted at the bottom of Fig 2. Increasingly accelerated worldsheets result in circles of lower and lower radius, until at $a\to\infty$, we get a point. This spacetime point is what is the D-instanton described above mathematically. The complementary picture is that when viewed from the $\zc$, $\za$ becomes a longer and longer string, gradually filling up all of spacetime to form a D-25 brane when the tension goes to zero \cite{Bagchi:2019cay}.

\smallskip
\noindent {\em{\underline{Reaching the Rindler Horizon II}}}. Finally, we discuss reaching the Rindler horizon at constant acceleration by evolving in time. We are interesting in Rindler time. So this is a limit $\eta\to \infty$. We will equivalently view this as 
\be{}
\eta \to \eta, \ \xi \to \e \xi, \, \e\to 0.
\ee
To understand this limit, we rewrite the 2d conformal generators in Rindler spacetime (we put $a=1$): 
\be{}
\L_n, \bL_n = \pm\frac{ i^n }{2} e^{n (\xi - \eta)}(\p_\eta \mp \p_\xi). 
\ee 
In the limit $\e\to0$, we get
\bea{}
 L_n &=& \L_n - \bL_{-n} = i^n e^{-n \eta}(\p_\eta - n \xi \p_\xi), \cr
 M_n &=& \e(\L_n + \bL_{-n}) = - i^n e^{-n \eta} \p_\xi
\eea
This closes to form the classical part of the BMS algebra \refb{bms} (i.e. $c_L=c_M=0$), as expected. This is again thus the null string, which we had expected. A detailed analysis about aspects of Rindler physics on constant accelerated worldsheets would be presented elsewhere \cite{Bagchi:upcoming}. 

\begin{figure}[t]
\centering
  \includegraphics[width=80mm]{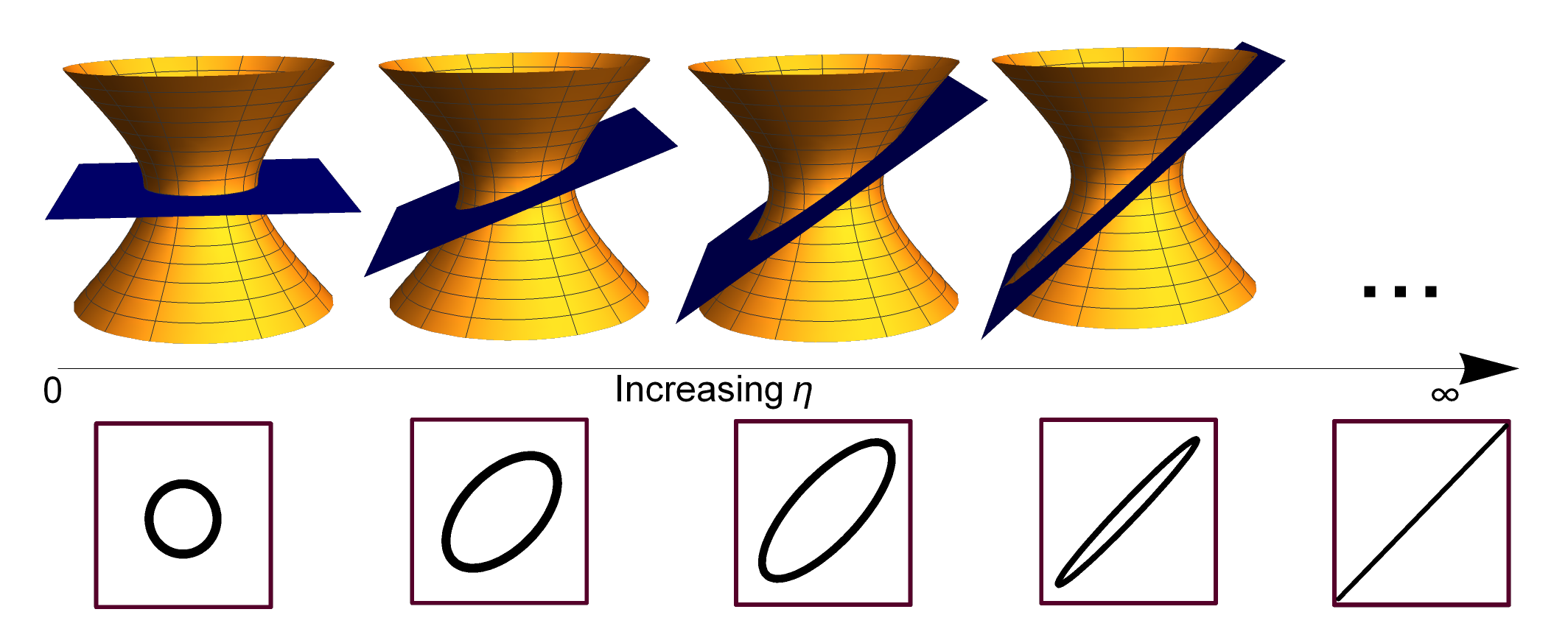}
  \caption{Equal time-slices of a Rindler worldsheet}
  \label{fig3}
\end{figure}

We now present a particularly intriguing picture that arises naturally on constant acceleration worldsheets. Notice that in Rindler spacetime, as depicted in Fig \ref{fig1}, constant Rindler time $(\eta)$ slices are straight lines through the origin with increasing slope on the Rindler R-wedge, depicted by $\eta_i$ ($\eta_1<\eta_2<\eta_3$). At $\eta\to \infty$, this hits the lightcone. On the L-wedge however, time runs backwards and the same slices are obtained by continuing R-wedge lines backwards into the third quadrant. For the string worldsheet at constant acceleration, the analogous picture is Fig \ref{fig3}. Increasing $\eta$ planes intersects the constant hyperboloid at increasing angles. The $\eta$-evolution of the closed string is shown in the boxes below. The circular closed string at the initial $\eta=0$ slice gets deformed as $\eta$ evolves. The tension decreases, the string gets longer and longer as given by the ellipses of increasing eccentricity. Ultimately, when $\eta\to \infty$, the lightcone is hit, the cross-section becomes a straight-line (an ellipse with eccentricity $=1$). The BMS algebra appears on the worldsheet as shown above. The string becomes tensionless and transitions into an open string from a closed one. 

\smallskip
\noindent {\em{\underline{Conclusions}}}. 
In summary, we have shown that Rindler physics on the worldsheet of the string captures many intriguing aspects of the tensionless string, including the emergence of open strings from closed strings and the remarkable null string complementarity. Our results would be particularly interesting for studies of strings near black holes where the near-horizon Rindler spacetime would induce a Rindler structure on the worldsheet. It has been speculated that strings grow long and hence become tensionless near black hole horizons \cite{Susskind:1993aa}. We have provided the tools for understanding this phenomenon from the worldsheet in this work. Details on this, as well as connections to thermal physics and entanglement on the worldsheet, would be elaborated on in \cite{Bagchi:upcoming}. Finally, the Carrollian limit \refb{vir2bms} of the Virasoro algebra and the Galilean limit result in the same BMS algebra \cite{Bagchi:2010eg}. From the string theory perspective, this hints at a duality between the extreme stringy limit and the supergravity regime of string theory. There are some interesting hints of this Galilean-Carrollian duality even in the Rindler story we have initiated here and we will expand on this in \cite{Bagchi:upcoming}. It will also be fascinating to link this to Torsional Newton-Cartan string theory \cite{Harmark:2017rpg, Harmark:2018cdl}, which exhibits the same algebra on the worldsheet.  

\smallskip
\noindent {\em{\underline{Acknowledgements}}}. This project has been a long time in the making, and we are grateful to a number of places for hospitality over the last couple of years especially Tsinghua University, Beijing and ULB Brussels, where the two ABs spent interesting times discussing and debating points that have culminated in this work. 

We are grateful to Tarek Anous, Rudranil Basu, Diptarka Das, Daniel Grumiller, Jelle Hartong, Gautam Mandal, Alejandro Rosabal and especially Joan Simon for helpful discussions and comments. We thank Ritankar Chatterjee and Pulastya Parekh for collaboration on related work. 

A Bagchi is partially supported by a Swarnajayanti fellowship from the Department of Science and Technology and the Science and Engineering Research Board (SERB) India, and other SERB grants: EMR/2016/008037, ERC/2017/000873, MTR/2017/000740. A Banerjee is supported by the Korea Ministry of Education, Science and Technology, ICT and Future Planning, Gyeongsangbuk-do and Pohang City. SC is partially supported by the ISIRD grant 9-252/2016/IITRPR/708.

\bibliography{ref}


\end{document}